\documentclass[twocolumn,showpacs,preprintnumbers,amsmath,amssymb]{revtex4}
\usepackage{color}
\usepackage{bm, graphicx, amsmath}
\usepackage{hyperref}

\begin{document}

\title{Coherence as a resource in decision problems:  The Deutsch-Jozsa algorithm and a variation}
\author{Mark Hillery$^{1,2}$}
\affiliation{$^{1}$Department of Physics, Hunter College of the City University of New York, 695 Park Avenue, New York, NY 10065 USA \\ $^{2}$Physics Program, Graduate Center of the City University of New York, 365 Fifth Avenue, New York, NY 10016}

\begin{abstract}
That superpositions of states can be useful for performing tasks in quantum systems has been known since the early days of quantum information, but only recently has a quantitative theory of quantum coherence been proposed.  Here we apply that theory to an analysis of the Deutsch-Jozsa algorithm, which depends on quantum coherence for its operation.  The Deutsch-Jozsa algorithm solves a decision problem, and we focus on a probabilistic version of that problem, comparing probability of being correct for both classical and quantum procedures.  In addition, we study a related decision problem in which the quantum procedure has one-sided error while the classical procedure has two-sided error.  The role of coherence on the quantum success probabilities in both of these problems is examined.
\end{abstract}  


\maketitle

\section{Introduction}
It is well known that entanglement is a resource that can be used for a number of tasks, for example, teleporting a quantum state from one system to another.  More recently, other quantum properties have been explored as resources.  The most recent is coherence \cite{baumgratz}.  Coherence is a basis-dependent property, and it depends on the off-diagonal matrix elements of the density matrix expressed in that basis.  The standard example is that of a particle going through an interferometer.  In order to see an interference pattern at the output, there has to be coherence between the paths the particle can take inside the interferometer.  One way to decrease the coherence between the paths is to gain information about which path the particle took, and doing so decreases the visibility of the interference pattern \cite{greenberger}-\cite{pati}. In \cite{baumgratz} two different ways of quantifying coherence were proposed, and we shall make use of one of them.

In order to treat a property, such as entanglement or coherence, as a resource, one needs a measure in order to quantify how much of that resource one has.  In the case of a pure, bipartite entangled state, the von Neumann entropy of one of the reduced density matrices of the state has proven to be a useful measure.  In the case of coherence, one defines a set of incoherent states (this set is basis-dependent), and the coherence of a state can be characterized by its distance from this set.  In \cite{baumgratz}, several possible distances were explored, and two with particularly nice properties were singled out.  One is based on relative entropy, and the other on the $l_{1}$ norm of the density matrix.  Here we shall use the latter.

In the context of coherence as a resource, it is useful to  see how the performance of a quantum algorithm that depends on coherence changes as the amount of coherence in the system decreases.  One of the first quantum algorithms, the Deutsch-Jozsa algorithm, depends on quantum coherence for its operation, and it is particularly simple \cite{deutsch}.  In fact, it can be rephrased as a particle going through a multi-arm interferometer and looking at the interference pattern at the output.  We will use a quantum walk version of the Deutsch-Josza algorithm to show this.  The Deutsch-Josza algorithm solves a decision problem and does the following.  One is given an oracle that evaluates a Boolean function, which is promised to be constant or balanced, and ones task is to determine which.  We will assume that our Boolean function maps $n$-bit strings to either $0$ or $1$, and if the input to the oracle is the string $x$, its output is $f(x)$.  A constant function is the same on all inputs and a balanced one is $0$ on half of the inputs and $1$ on the others.  In the worst case scenario, one would have to check $2^{n-1}+1$ inputs to be certain which kind of function one had, while in the quantum case only one function evaluation is necessary.  

If one is willing to accept a probabilistic answer, classically one would only have to check a few inputs in order to determine which type of function the oracle represented with a small probability of making a mistake.  Consequently, the Deutsch-Jozsa algorithm is not a practical one, but it does serve to illustrate how quantum mechanics allows one to perform tasks in a different way than would be possible on a classical computer, and gain some quantum advantage.  The classical-quantum comparison can be made precise by asking for the probability of obtaining the correct answer, constant or balanced, in a fixed number of runs.

Here we wish to examine the effect of decoherence on the performance of the Deutsch-Jozsa algorithm, and a variant of it, using a recently defined measure of coherence \cite{baumgratz}.  The Deuthsch-Jozsa algorithm depends on quantum coherence, and the less of it there is, the worse the algorithm will perform.  We wish to make this statement quantitative using one of the measures for coherence proposed in \cite{baumgratz} and several different measures for the performance of the algorithm.  We will see how the amount of coherence affects our ability to distinguish the balanced and constant cases for a fixed number of measurements and compare this to the result of a classical procedure.  We will then examine modified decision problem, deciding between a balanced function and one that is biased, i.e.\ $\sum_{x}f(x) = \epsilon 2^{n}$, where $\epsilon$ is known.

\section{Quantum walk}
We will use a quantum walk version of the Deutsch-Jozsa algorithm.  The reason for doing so is that this version of the algorithm shows that the Deutsch-Jozsa algorithm is analogous to sending a particle through an interferometer that has a large number of paths.  This use of the interferometer makes it clear that the quantum resource we are using is just quantum coherence.  The graph on which the walk takes place is shown in Figure 1.  The tails on the graph are semi-infinite, with the right-hand tail having vertices $N+1$, $N+2$, \ldots, and the left-hand tail having vertices $0$, $-1$, $-2$, and so on.
\begin{figure}[h!]
\centering
\includegraphics[width=0.5\textwidth]{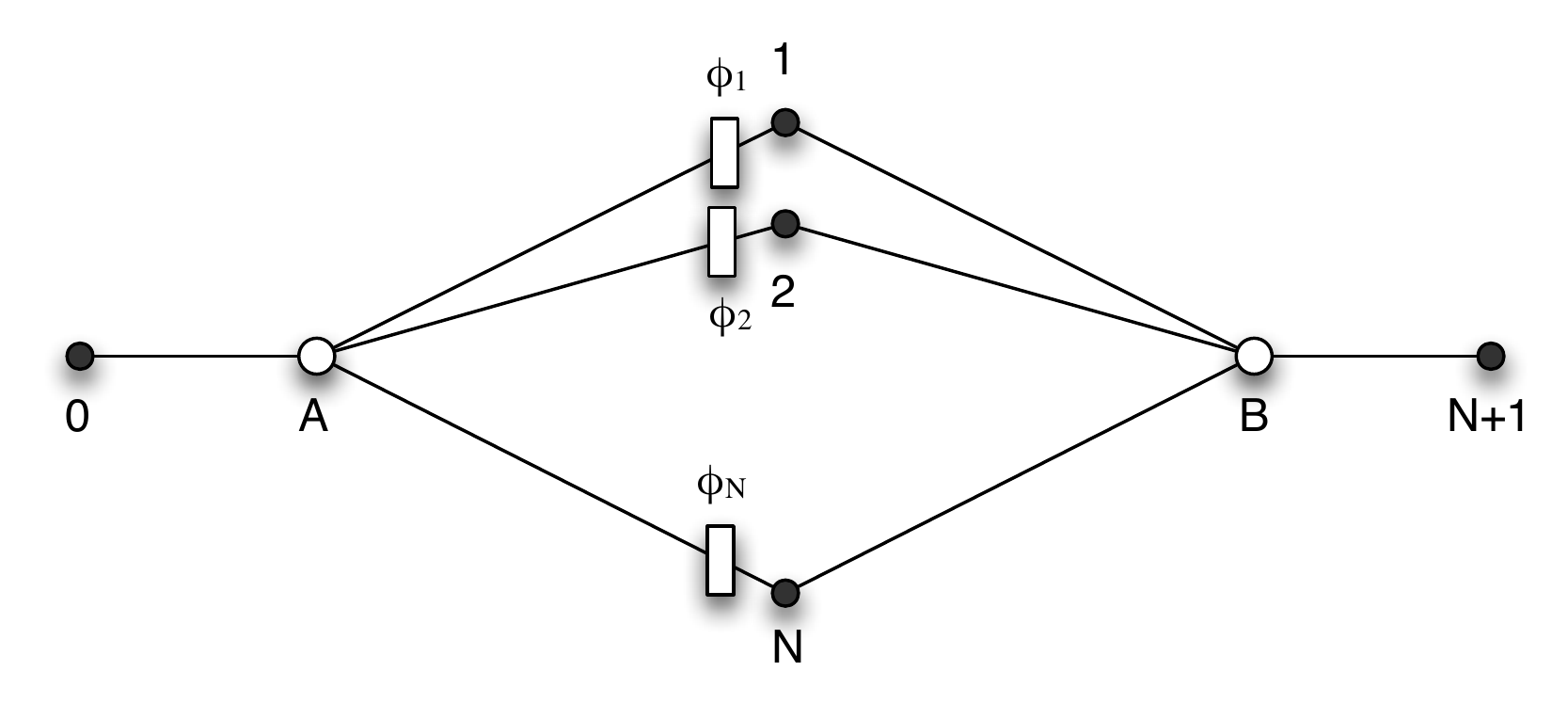} 
\caption{Graph on which the quantum walk takes place.  The vertices $A$ and $B$ are Fourier vertices, and all other vertices simply transmit the particle.  There are $N$ paths going from vertex $A$ to vertex $B$.  The rectangles are phase shifters and the $j^{\rm th}$ one multiplies the state by $\exp (i\phi_{j})$.  The tails, one starting at the vertex $0$ and the other starting at the vertex $N+1$, are semi-infinite.}
\label{graph}
\end{figure}
We will be using a scattering walk, which is a discrete-time quantum walk \cite{hillery1}.  In this type of walk, the particle sits on the edges, not the vertices, and each edge has two orthogonal states, each corresponding to the particle moving in a particular direction.  For example, the edge between $0$ and $A$ has the states $|0,A\rangle$ corresponding to the particle being on that edge and moving from $0$ to $A$, and the state $|A,0\rangle$ corresponding to the particle being on that edge and moving from $A$ to $0$.  To each vertex of the graph corresponds a unitary operator that transforms states entering the vertex into states leaving the vertex.  The unitary operator that advances the walk one time step is composed of the combined actions of the unitary operators at the individual vertices.  The vertices $A$ and $B$ are Fourier transform vertices and the unitary operators corresponding to them, $U_{A}$ and $U_{B}$ respectively, act as
\begin{eqnarray}
U_{A}|j,A\rangle & = & \frac{1}{\sqrt{N+1}}\sum_{k=0}^{N} e^{2\pi jk/(N+1)}|A,k\rangle \nonumber \\
U_{B}|j,B\rangle & = & \frac{1}{\sqrt{N+1}}\sum_{k=1}^{N+1} e^{2\pi jk/(N+1)}|B,k\rangle .
\end{eqnarray}
The other vertices just transmit the particle, but those with a phase shifter also add a phase factor to the transmitted state,
\begin{equation}
|A,j\rangle \rightarrow e^{i\phi_{j}} |j,B\rangle .
\end{equation}
In our case the phases, $\phi_{j}$, will be either $0$ or $\pi$, and these phases correspond to the output of the Boolean function in the Deutsch-Jozsa algorithm.  The phases are promised either to be all the same (constant) or half of them are $0$ and half are $\pi$ (balanced).  Our task is to find out which of the two cases we have.  We will start the particle in the state $|0,A\rangle$, run the walk for 3 steps, and then see whether or not it is in the state $|B,N+1\rangle$.  If we find the particle in that state we will conclude the phases were all the same, and if we do not, we will conclude we had the balanced situation.

In order to compare the quantum walk result to a classical one, we will assume that classically we are able to sample the phase shifters, i.e. pick some of them and see how they are set, whether to $0$ or $\pi$.  Then a classical versus quantum comparison will consist of a comparison between the number of phase shifters we sample versus the number of times we have to run the quantum walk.

\section{Analysis of the walk}
We start the particle making the walk in the state $|0,A\rangle$.  After two steps, its state is 
\begin{equation}
\frac{1}{\sqrt{N+1}} \left[ |0,-1\rangle + \sum_{j=1}^{N}e^{i\phi_{j}}|j,B\rangle \right] .
\end{equation}
One more step yields the state
\begin{eqnarray}
\frac{1}{\sqrt{N+1}}|-1,-2\rangle \nonumber \\
+ \frac{1}{N+1}\sum_{j=1}^{N}\sum_{k=1}^{N}e^{i\phi_{j}}
e^{2\pi i jk/(N+1)} |B,k\rangle \nonumber \\
+\frac{1}{N+1}\left( \sum_{j=1}^{N} e^{i\phi_{j}}\right) |B,N+1\rangle .
\end{eqnarray}
The last term is the one that interests us, because it yields the probability that the particle is on the edge between $B$ and $N+1$.  If all of the phases, $\phi_{j}$ are the same, this probability is just $N^{2}/(N+1)^{2}$, and if half the phases are $0$ and half $\pi$, then, assuming $N$ is even, it is zero.  Therefore, with a small error of order $1/N$, which we shall assume we can neglect, we can determine which of these two possibilities we have by measuring the walk after three steps to see whether the particle is between $B$ and $N+1$ or not.

Now we want to introduce decoherence into this system.  One way of doing so is to introduce a qubit for each leg of the graph.  In particular, let us suppose that all of these qubits are initially in the state $|0\rangle$.  When the particle goes through the vertex $j$, in addition to picking up the phase $e^{i\phi_{j}}$, the $j^{\rm th}$ qubit goes from the state $|0\rangle_{j}$ to the state $|\mu_{j}\rangle_{j}$, which is a linear combination of the states $|0\rangle_{j}$ and $|1\rangle_{j}$, $|\mu_{j}\rangle_{j} = \alpha_{j} |0\rangle_{j} + \beta_{j} |1\rangle_{j}$.  If we let 
\begin{equation}
|\eta_{j}\rangle = |\mu_{j}\rangle_{j} \prod_{k=1.k\neq j}^{N}|0\rangle_{k} ,
\end{equation}
for $j=1,2,\ldots N$ and $|\eta_{0}\rangle = \prod_{k=1}^{N}|0\rangle_{k}$, then the state after two steps is
\begin{equation}
\label{2steps}
\frac{1}{\sqrt{N+1}} \left[ |0,-1\rangle |\eta_{0}\rangle + \sum_{j=1}^{N}e^{i\phi_{j}}|j,B\rangle |\eta_{j}\rangle \right] .
\end{equation}
The reduced density matrix corresponding to this state, where we trace out the ancilla qubits, is given by
\begin{eqnarray}
\rho_{int} & = & \frac{1}{N+1} \sum_{j=1}^{N}\sum_{k=1}^{N} e^{i(\phi_{j}-\phi_{k})}\langle\eta_{k}|\eta_{j}\rangle |j,B\rangle\langle k,B| \nonumber \\
& & + O(N^{-1/2}) .
\end{eqnarray}
One of the measures of coherence defined in \cite{baumgratz} for a general density matrix, $\rho$, on an $M$-dimensional space is
\begin{equation}
C_{l_{1}}(\rho )= \sum_{i,j=1, i\neq j}^{M} |\rho_{ij}| .
\end{equation}
If we define 
\begin{equation}
X=\frac{1}{(N+1)^{2}}\sum_{j,k=1,j\neq k}^{N} |\langle \eta_{k}|\eta_{j}\rangle | ,
\end{equation}
then we see that 
\begin{equation}
C_{l_{1}}(\rho_{int} )=(N+1) X + O(N^{-1/2}) .
\end{equation}
Note that in the case in which all of the qubit states $|\mu_{j}\rangle_{j}$ are the same, then the inner products $\langle \eta_{k}|\eta_{j}\rangle$ are independent of $j$ and $k$ for the case $j\neq k$.  Setting $\nu = | \langle \eta_{k}|\eta_{j}\rangle |$, we then find that $X=\nu N(N-1)/(N+1)^{2}$.  If we now let the walk go one more step, the state is 
\begin{eqnarray}
\frac{1}{\sqrt{N+1}}|-1,-2\rangle |\eta_{0}\rangle \nonumber \\
+ \frac{1}{N+1}\sum_{j=1}^{N}\sum_{k=1}^{N}e^{i\phi_{j}} e^{2\pi i jk/(N+1)} |B,k\rangle |\eta_{j}\rangle \nonumber \\
+\frac{1}{N+1}\left( \sum_{j=1}^{N} e^{i\phi_{j}} |\eta_{j}\rangle \right) |B,N+1\rangle  .
\end{eqnarray}
Forming a density matrix from this state and tracing out the ancillas gives us the output density matrix for the particle making the walk, $\rho_{out}$, and the probability of finding the particle on the edge between $B$ and $N+1$ is 
\begin{eqnarray}
\langle B, N+1|\rho_{out}|B, N+1\rangle =  \frac{1}{(N+1)^{2}} \nonumber \\
\left( \sum_{j=1}^{N}\sum_{k=1}^{N} e^{i(\phi_{j}-\phi_{k})}\langle \eta_{k}|\eta_{j}\rangle\right)  .
\end{eqnarray}
Note that
\begin{equation}
\langle B, N+1|\rho_{out}|B, N+1\rangle  \leq \frac{N}{(N+1)^{2}} + X.
\end{equation}
What this tells is is that the amount of coherence in the system places an upper limit on our ability distinguish the constant and balanced cases.  With perfect coherence the particle always (up to $O(1/N)$) finishes in the state $|B,N+1\rangle$ and in the balanced case it never does.  When the amount of coherence decreases, the the probability that the constant case will be mistaken for the balanced case increases.  This shows that the quantum resource that is being used to accomplish this task is coherence.

\section{Deutsch-Jozsa algorithm}

Now let us see what happens to the results of the Deutsch-Jozsa algorithm as the amount of coherence in the system is decreased in more detail.  In particular, we will examine the probability of correctly identifying whether the interferometer is constant or balanced in a fixed number of runs.  We shall look at the case that $\langle\eta_{k}|\eta_{j}\rangle$ is independent of $j$ and $k$ for $j\neq k$ and we shall assume the inner product is real and positive so we can set $\langle\eta_{k}|\eta_{j}\rangle=|\langle\eta_{k}|\eta_{j}\rangle |=\nu$.  If all of the $\phi_{j}$ are the same, then 
\begin{eqnarray}
\langle B, N+1|\rho_{out}|B, N+1\rangle & = & \frac{1}{(N+1)^{2}} [ N + \nu N(N-1)] \nonumber \\
 & \simeq & \nu +O(1/N) .
\end{eqnarray}
If half of the $\phi_{j}$ are $0$ and half are $\pi$, then we have that
\begin{eqnarray}
\langle B, N+1|\rho_{out}|B, N+1\rangle & =  & \frac{1}{(N+1)^{2}} \left[ (1-\nu )N \right. \nonumber \\
& &\left.  + \nu \sum_{j,k=1}^{N} e^{i(\phi_{j}-\phi_{k})} \right]  \nonumber \\
 & = & \frac{(1-\nu )N}{(N+1)^{2}} \nonumber \\
 & =&  O(1/N) .
\end{eqnarray}
Our procedure is to run the walk and measure whether the particle is in the state $|B,N+1\rangle$.  If it is, we guess that we have the constant case, and if not, we guess we have the balanced case.
We see that as the amount of coherence decreases, our chance of making an error increases.  Note that the error is almost one-sided.  If the particle comes out, we know with very high probability that all of the $\phi_{j}$ are the same.  However, if it does not come out, and we guess that the $\phi_{j}$ are in the balanced configuration, then, assuming the balanced and constant cases are equally likely, we have a chance of $(1-\nu )/2$ of being wrong.  Classically, looking at one of the phase shifters gives us no information about which of the two cases we have, so for one trial, the quantum case does better.  Clearly, coherence is a resource in the quantum case, because the more coherence there is in the system, the less likely we are to make a mistake.

Now let's see what happens with two trials.  Let's look at the classical case first.  We shall call the results of the trials $y_{1}$ and $y_{2}$, where $y_{j}=\pm 1$.  Here we are denoting the phase shifters by $e^{i\phi}$ rather than $\phi$, so a phase of $0$ corresponds to $1$ and a phase of $\pi$ corresponds to $-1$.  There are four possible results, $(y_{1},y_{2})$, if we sample two of the phase shifters, $(1,1)$, $(1,-1)$, $(-1,1)$, and $(-1,-1)$.   We will assume that the balanced and constant cases are equally likely, and that within the constant category, each value is equally likely.  If the results are different we know we have the balanced case.  This happens with a probability of $1/4$ (the probability of the balanced case occurring times the probability of the results being different).  If we get the same result for each trial, things get a bit more complicated.  We want to find $P(c|y_{1}y_{2})$, the probability that we have the constant case given that we have the result $(y_{1},y_{2})$, and similarly $P(b|y_{1},y_{2})$, the probability that we have the balanced case.  Clearly $P(b|1,-1)=P(b|-1,1)=1$.  To find the probabilities when the results are the same, we use Bayes' theorem.  Let us find $P(c|1,1)=P(c=1|1,1)$, where we have specifically indicated which constant value the phase shifter will have.  We then have
\begin{equation}
P(c=1|1,1)=\frac{P(1,1|c=1)P(c=1)}{P(1,1)} .
\end{equation}
Now $P(1,1|c=1) = 1$ and $P(c=1)=1/4$.  For the denominator we have
\begin{eqnarray}
P(1,1) & = & P(1,1|c=1)P(c=1) + P(1,1|b)P(b) \nonumber \\
& = & \frac{3}{8} .
\end{eqnarray}
Finally, this gives us that $P(c|1,1)=2/3$, which implies that $P(b|1,1)=1/3$.  Similarly, $P(c|-1,-1)=2/3$.  Our strategy, then, is to guess balanced if the results are different, and constant, if they are the same.  Our probability of being correct is $3/4$, i.e.\ we are always correct if the results are different and are correct with a probability $2/3$ when they are the same.

Now let us look at the quantum case.  We run the walk twice, and we denote the results of the runs by $(0,0)$, $(1,0)$, $(0,1)$ and $(1,1)$, where $0$ denotes we did not find the particle in the state $|B, N+1\rangle$ and $1$ indicates that we did.  Neglecting terms of $O(1/N)$, we have that
\begin{eqnarray}
P(0,0|c) & = & (1-\nu )^{2} \nonumber \\ 
P(0,1|c) & = & P(1,0|c) =\nu (1-\nu)  \nonumber \\
P(1,1|c) & = & \nu^{2}  ,
\end{eqnarray}
and $P(0,0|b)=1$ and $P(0,1|b)=P(1,0|b)=P(1,1|b)=0$.  Now making use of Bayes' theorem we have that in the cases $(1,0)$, $(0,1)$ and $(1,1)$, we can conclude that we have the constant case with certainty.  If we obtain $(0,0)$, then we have
\begin{eqnarray}
P(c|0,0) & = & \frac{(1-\nu )^{2}}{(1-\nu )^{2}+1} \nonumber \\
P(b|0,0) & = & \frac{1}{(1-\nu)^{2}+1} .
\end{eqnarray}
Now the first of these probabilities is less than or equal to the second, so if our measurement results are $(0,0)$, we should always guess balanced.  In all other cases we guess constant.  Doing so, our probability of being wrong is $(1/2)(1-\nu )^{2}$.  The quantum error probability will be less than the classical one when
\begin{equation}
\frac{1}{2}(1-\nu )^{2} < \frac{1}{4} ,
\end{equation}
or $\nu > (\sqrt{2}-1)/\sqrt{2}$.  So in the case of two trials, as long as the amount of decoherence is not too great, the quantum method is better.

This can be generalized to $m$ trials for $m\ll N$.  Before doing so, let us be more careful about specifying our ensemble.  We are assuming that each of the constant cases occurs with probability $1/4$, and that the total probability of the balanced case is $1/2$.  Within the balanced case, each of the balanced sequences has the same probability.  So far, we have assumed that, given the balanced case, this is equivalent to the probability that a particular phase shifter has $y_{j}=1$ is $1/2$, the probability that it has $y_{j}=-1$ is $1/2$, and that different phase shifters can be treated as independent.  This needs to be justified, and this is done in the Appendix.  We find that as long as $m\ll N$, this assumption is valid.

In the classical case, the only ambiguous situation is if all of the examined phase shifters are found to be the same.  We would then guess that we are in the constant situation.  In the quantum case, the only ambiguous case is if the particle is never found between $B$ and $N+1$. We would then guess that we are in the balanced situation.  Let us have a look at these cases and see what the probability of making a mistake is.  In all other situations, the probability of making a mistake is very small.

We start with the classical case.  Denote the probability that we have $c=1$ given that we examined $m$ phase shifters and found them to be $1$ by $P(c=1|m1)$.  Making use of Bayes' theorem and 
\begin{eqnarray}
P(m1) & = & P(m1|c=1)P(c=1)+P(m1|b)P(b) \nonumber \\
& = & \frac{1}{4}+\frac{1}{2^{m+1}} ,
\end{eqnarray}
we find that 
\begin{equation}
P(c=1|m1)= \frac{2^{m-1}}{1+2^{m-1}} ,
\end{equation}
The result for the probability for $c=-1$ when we found $m$ phase shifters to be $-1$ is the same.  Since we will guess the constant case in both these situations, the probability of being wrong is
\begin{eqnarray}
p_{error}^{(class)} & =&  2[1-P(c=1|m1)]P(m1) \nonumber \\
& = & \frac{2}{1+2^{m-1}}\left( \frac{1}{4}+\frac{1}{2^{m+1}} \right) = \frac{1}{2^{m}}.
\end{eqnarray}
Now we move to the quantum case, and $P(c|m0)$ now denotes the probability that we have the constant case given that the particle was not found in the state $|B,N+1\rangle$ in $m$ trials.  Now application of Bayes' theorem and the fact that
\begin{eqnarray}
P(m0) & = & P(m0|c)P(c)+P(m0|b)P(b) \nonumber \\
& = &  \frac{1}{2}(1-\nu )^{m}+\frac{1}{2} 
\end{eqnarray}
gives us
\begin{equation}
P(c|m0)=\frac{(1-\nu )^{m}}{1+(1-\nu )^{m}} .
\end{equation}
Now in this case we will guess balanced, so the probability of being wrong is
\begin{equation}
p_{error}^{(quant)}=P(c|m0)P(m0)=\frac{1}{2}(1-\nu )^{m} .
\end{equation}
If $\nu > 1-(2^{1/m}/2)$, then we will have $p_{error}^{(quant)}< p_{error}^{(class)}$.  This tells us how much coherence we need for the quantum method to outperform the classical one.

\section{Variation on Deutsch-Jozsa}
The decision problem we looked at in the previous section was one in which both the quantum algorithm and the classical one had (almost) one-sided error.  In the classical case, if the interferometer is constant, we will never guess balanced, and in the quantum case, if it is balanced, we will never guess constant.  Here we would like to look at a situation in which the quantum algorithm has one-sided error, but the classical one does not.  This can give the quantum algorithm a significant advantage if the errors have different costs.  We again look at the case where the phase shifts are either $0$ or $\pi$, but we now want to distinguish between the case in which the phase shifts are balanced and the case in which 
\begin{equation}
\frac{1}{N}\sum_{j=1}^{N}e^{i\phi_{j}}=\epsilon  ,
\end{equation}
where we assume that $\epsilon \ll 1$.  In order to distinguish between these alternatives, our strategies are the same as before.  The quantum strategy is to run a quantum walk a certain number of times, and the classical strategy is to sample the phase shifters.

In this case, the quantum strategy is the easier one to analyze.  Let us first consider the situation without decoherence.  We know that in the balanced case, the probability of measuring the particle to be in the state $|B,N+1\rangle$ after the walk is, up to $O(1/N)$, zero.  In the second case, which we shall refer to as the $\epsilon$ case, the probability to find the particle in that state is $\epsilon^{2}$.  In that case, if the walk is run $m$ times, the probability of not finding the particle between $B$ and $N+1$ is
\begin{equation}
(1-\epsilon^{2})^{m} \simeq e^{-m\epsilon^{2}} ,
\end{equation}
where we have made use of the fact that $\ln (1-\epsilon^{2})^{m} \simeq -m\epsilon^{2}$.  Therefore, in order to detect this case, that is to find the particle at least once in the state $|B,N+1\rangle$, we need $m\epsilon^{2}$ to be at least of order $1$.  Our strategy is to assume that if we ever find the particle in the state $|B,N+1\rangle$ in $m$ runs that we have the $\epsilon$ case, and that we have the balanced case otherwise.  If we are given the balanced case, we will always be correct, and if we are given the $\epsilon$ case and $m\epsilon^{2}$ is of order $1$ or greater, our probability of error will also be small. If there is decoherence, the effect is simply to replace $\epsilon^{2}$ by $\nu\epsilon^{2}$, so that as long as $\nu$ is not too small, the effect of decoherence will not be large

Now we turn to the classical case.  We will look at $m$ of the phase shifters.  Let each sampled phase shifter be represented by a variable $y_{j}$, where $y_{j}=1$ corresponds to $\phi_{j}=0$ and $y_{j}=-1$ corresponds to $\phi_{j}=\pi$.  We define
\begin{equation}
Y=\frac{1}{m}\sum_{j=1}^{m}y_{j} .
\end{equation}
If we find $Y\geq \epsilon /2$ we shall assume that we have the $\epsilon$ case, otherwise we will assume we have the balanced case.  Therefore, we want to find the probability of making an error.

Let us start by assuming that we have the balanced case, and we would like to find the probability that we would identify it as the $\epsilon$ case.  We will assume that all of the balanced sequences of $N$ phase shifters are equally probable.  If we are only sampling $m\ll N$ of the phase shifters, this is equivalent to assuming that each phase shifter we look at has an equal chance of having $y_{j}=1$ and $y_{j}=-1$ (see Appendix).  We now want to find the probability that $Y\geq \epsilon /2$.  For this purpose we can use the Chernoff bound \cite{motwani}.  It states that if we have the independent random variables, $X_{j}$,  $j=1,2,\ldots n$, where $X_{j}$ can be either $0$ or $1$, and its probability of being $1$ is $p_{j}$, then for $X_{T}=\sum_{j=1}^{n}X_{j}$, $\mu=\sum_{j=1}^{n}p_{j}$, and any $\delta >0$, then
\begin{equation}
P(X_{T}>(1+\delta )\mu ) < \left[ \frac{e^{\delta}}{(1+\delta )^{(1+\delta )}} \right]^{\mu} .
\end{equation}
In our case $\mu = m/2$, and setting $X_{j}=(1/2)(y_{j}+1)$, we find that $Y>\epsilon /2$ implies $X_{T}>m[(\epsilon /2) + 1)/2$ so that
\begin{equation}
\label{errorprob}
P(Y>\epsilon /2 )<\left[\frac{e^{\epsilon /2}}{(1+\epsilon /2 )^{1+\epsilon /2}} \right]^{m/2} .
\end{equation}
Assuming $\epsilon \ll 1$ and keeping lowest order terms in $\epsilon$, we find 
\begin{equation}
\ln \left[\frac{e^{\epsilon /2}}{(1+\epsilon /2)^{1+\epsilon /2}} \right] \simeq -\epsilon^{2}/4 ,
\end{equation}
so that the right-hand side of Eq.\ (\ref{errorprob}) is approximately $e^{-\epsilon^{2}m/8}$.

Similarly, let us suppose that we have the $\epsilon$ case.  We will assume that all sequences of $N$ phase shifters satisfying $\sum_{j=1}^{N}y_{j}=\epsilon N$ are equally likely.  For a subsequence of length $m$, where $m\ll N$, this is equivalent to assuming that each element has a probability of $(1+\epsilon )/2$ to be $+1$ and $(1-\epsilon )/2$ to be $-1$ (see Appendix).   We now want to find the probability that we would identify this as the balanced case, which is the same as finding $Y<\epsilon /2$.  We can now use the following version of the Chernoff bound \cite{motwani}.  With the same conditions as before, 
\begin{equation}
P(X_{T}<(1-\delta )\mu ) < e^{-\mu \delta^{2}/2} .
\end{equation}
We now have that $\mu = m(1+\epsilon )/2$ and $Y< \epsilon /2$ implies $X_{T}< m[(\epsilon /2) + 1)/2$, which further implies that
\begin{equation}
\delta = \frac{\epsilon}{2(1+\epsilon)} .
\end{equation}
Finally, keeping only lowest order terms in $\epsilon$, we find that
\begin{equation}
P(Y<\epsilon /2 ) < e^{-\epsilon^{2}m/8} .
\end{equation}

Summarizing, we see the following.  For both the quantum and classical methods, the condition for keeping the error small is the same, $m\epsilon^{2}$ should be at least of order $1$.  However, up to $O(1/N)$, the quantum error is one-sided, if we have the balanced case, we will not mistake it for the $\epsilon$ case.  For the classical method, the error is two-sided, we can mistake each case for the other.  Therefore, if we are in a situation in which the cost of mistaking the balanced case for the $\epsilon$ case is large, the quantum method has an advantage.  Note that for this situation, deciding between the balanced and $\epsilon$ cases, the type of decoherence we are considering does not affect the fact that the quantum error is one sided, but it will cause the number of runs that we need to make in the quantum case, which is of order $1/(\nu\epsilon^{2})$, to increase.  The reason it does not affect the one-sidedness of the error, is that the decoherence respects the symmetry of the problem; it is the same for each branch of the interferometer.  This suggests that for some problems for which coherence is a resource, not only its total quantity, but its properties will play a role.

\section{Conclusion}
We have examined the role played by coherence as a resource in the Deutsch-Jozsa and related algorithms.  The Deutsch-Jozsa algorithm is a means of solving a decision problem, in particular, deciding between two alternatives.  In its ideal form, it provides an answer in a single run, whereas classically in the worst case an exponential number of runs would be necessary.  Decoherence degrades the ability of the algorithm to decide between the alternatives, and the smaller the amount of coherence in the system, the worse the ability of the algorithm to distinguish between the two cases.  This demonstrates that coherence is a resource for this algorithm.  We also looked at the Deutsch-Jozsa algorithm in a probabilistic setting, and found that as long as there is enough coherence present, there is a quantum advantage in that for a fixed number of measurements, one has a higher probability of making the correct decision using quantum means than by using classical ones.  By looking at a related decision problem, we found an example in which the number of measurements one makes is comparable for the classical and quantum cases, at least if the coherence in the quantum case remains high enough, but while the classical procedure has two-sided error, the quantum procedure has one sided error.

\section*{Acknowledgment}
This research was supported by a grant from the John Templeton Foundation.  I would like to thank Seth Cottrell, Emilio Bagan, and Janos Bergou for useful conversations.

\section*{Appendix}
We now need to justify what we did in Sections III and IV.  In the our ensemble in Section III, each balanced sequence of length $N$ occurred with equal probability.  A related ensemble occurred in Section IV.  We want to show the following.  We consider an ensemble of sequences of length $N$ consisting of $\pm 1$, in which each sequence has $pN$ ones and $(1-p)N$ minus ones.  Each of these sequences has the same probability.  We now consider fixed subsequences of these sequences of length $m$, e.g.\ the first $m$ elements of each sequence of length $N$.  We want to show that the probability of a subsequence with $m_{+}$ ones and $m_{-}$ minus ones, where $m_{+}+m_{-}=m$,  is the same as if each location in the subsequence has a probability $p$ of containing a one and a probability $1-p$ of containing a minus one.  

For convenience, we will consider subsequences consisting of the first $m$ places of the sequences of length $N$.  The probability, $p(m_{+},m_{-})$ that the subsequence has $m_{+}$ ones is
\begin{eqnarray}
p(m_{+},m_{-}) & = & \frac{\left(\begin{array}{c}m \\ m_{+} \end{array}\right)  \left(\begin{array}{c}N-m \\ pN-m_{+} \end{array}\right)} {\left(\begin{array}{c}N \\ pN \end{array}\right)} \nonumber \\
 & = & \left(\begin{array}{c}m \\ m_{+} \end{array}\right) F,
 \end{eqnarray}
where
\begin{equation}
F=\frac{(N-m)! (pN)! (N-pN)!}{N! (pN-m_{+})! (N-pN-m+m_{+})!} .
\end{equation}
Now we shall assume that $m$ is much less than $N$, $pN$, and $(1-p)N$ and apply the Stirling approximation, $n!\simeq \sqrt{2\pi n} n^{n}e^{-n}$.  We then have 
\begin{equation}
\frac{(N-m)!}{N!} \simeq \sqrt{1-\frac{m}{N}} \frac{e^{m}}{N^{m}}\left( 1-\frac{m}{N}\right)^{N-m} .
\end{equation}
We can approximate the last factor by taking its logarithm and expanding in $m/N$
\begin{eqnarray}
\ln \left( 1-\frac{m}{N}\right)^{N-m} & = & (N-m)\left( -\frac{m}{N}-\frac{m^{2}}{2N} + \dots\right) \nonumber \\
& = & -m-\frac{m^{2}}{2N} +O(m/N) .
\end{eqnarray}
This gives us
\begin{equation}
\frac{(N-m)!}{N!} \simeq \sqrt{1-\frac{m}{N}} \frac{1}{N^{m}}e^{-m^{2}/2N} .
\end{equation}
Applying this relation to $(pN-m_{+})!/(pN)!$ and $[(1-p)N-(m-m_{+})]!/[(1-p)N]!$ and substituting into the expression for $F$, we obtain
\begin{equation}
F\simeq p^{m_{+}}(1-p)^{m-m_{+}} + O(m/\sqrt{N}) .
\end{equation}
This gives us
\begin{equation}
p(m_{+},m_{-}) \simeq \left(\begin{array}{c}m \\ m_{+} \end{array}\right) p^{m_{+}}(1-p)^{m-m_{+}} ,
\end{equation}
which is what we would obtain if we assumed that in the sequence of length $m$ one occurred with probability $p$ and minus one occurred with probability $1-p$.


\begin{thebibliography}{99}
\bibitem{baumgratz} T.\ Baumgratz, M.\ Cramer, and M.\ B.\ Plenio, Phys.\ Rev.\ Lett.\ {\bf 113}, 140401 (2014).
\bibitem{greenberger} D.\ M.\ Greenberger and A.\ YaSin, Phys.\ Lett.\ A {\bf 128}, 391 (1988).
\bibitem{jaeger} G.\ Jaeger, A. Shimony, and L.\ Vaidman, Phys.\ Rev.\ A {\bf 51}, 54 (1995).
\bibitem{englert} B-.\ G.\ Englert, Phys.\ Rev.\ Lett.\ {\bf 77}, 2154 (1996).
\bibitem{durr} S.\ D\"{u}rr, Phys.\ Rev.\ A {\bf 64}, 042113 (2001).
\bibitem{bergou1} B-.\ G.\ Englert and J.\ Bergou, Opt.\ Commun.\ {\bf 179}, 337 (2000).
\bibitem{bergou2} M.\ Jakob and J.\ A.\ Bergou, Phys.\ Rev.\ A {\bf 76}, 052107 (2007).
\bibitem{pati} M.\ N.\ Bera, T.\ Qureshi, M.\ A.\ Siddiqui, and A.\ K.\ Pati, Phys.\ Rev.\ A {\bf 92}, 012118 (2015).
\bibitem{deutsch} D.\ Deutsch and R.\ Jozsa, Proc.\ Royal Doc.\ London A {\bf 439},  553 (1992). 
\bibitem{hillery1}M.\ Hillery, J.\ Bergou, and E.\ Feldman, Phys.\ Rev.\ A {\bf 68}, 032314 (2003).
\bibitem{motwani} R.\ Motwani and P.\ Raghavan, \emph{Randomized Algorithms}, (Cambridge University Presss, Cambridge, 1995).
\end{thebibliography}
\end{document}